\documentclass[aps,pre,twocolumn,floatfix]{revtex4}

\usepackage{graphicx}
\usepackage{dcolumn}
\usepackage{bm}

\begin{document}


\title{Steinberg-Guinan strength model for rhenium}

\date{July 24, 2025 -- LLNL-TR-2009087} 

\author{Damian C. Swift}
\email{dswift@llnl.gov}
\affiliation{%
   Physics Division, Lawrence Livermore National Laboratory,
   7000 East Avenue, Livermore, California 94551, USA
}

\begin{abstract}
Rhenium, Re, is used as an x-ray shield in laser-driven material property experiments,
where its strength at high pressures can be a consideration in the design, modeling, and interpretation.
We present a Steinberg-Guinan (SG) strength model for Re,
tailored for use in high-pressure dynamic loading simulations.
Parameters for the SG model were derived from recent atom-in-jellium predictions of the shear modulus under compression
and experimental data on work-hardening from rolled-bar studies.
The ambient shear modulus was fixed to the measured value,
and the pressure-hardening parameter was fitted to the atom-in-jellium predictions up to 1\,TPa.
The shear modulus model was still a reasonable fit beyond 25\,TPa.
Thermal softening was estimated from the thermal expansivity and bulk modulus.
Work-hardening parameters were extracted by fitting the model to Knoop microhardness measurements under known plastic strains.
The resulting model captures the observed hardening behavior but predicts significantly lower flow stresses at high pressures than diamond anvil cell observations suggest, implying that Re may exhibit enhanced strength at megabar pressures.
These results provide a basis for improved modeling of strength in Re under extreme conditions and suggest directions for further theoretical and experimental investigation.
\end{abstract}


\maketitle

\section{Introduction}
In dynamic loading experiments
on the properties of materials at the highest pressures,
the load is generated by laser-induced ablation.
Hard x-rays emitted by the plasma formed from the ablator 
penetrate the target assembly, causing generally unwanted heating.
Usually, a layer of high-$Z$ elements is used to absorb these x-rays
at the back of the ablator.
The element most commonly used has been Au.
For Be ablators, which are used commonly in experiments studying the
response of materials to dynamic loading,
it is advantageous to also include Re, 
as the combined absorption spectrum is substantially more effective in
absorbing kilovolt-range x-rays emitted by the Be plasma.
The shield layers are usually $\sim$1\,$\mu$m thick.

In x-ray diffraction experiments, we usually try to ensure that these x-ray shields melt.
If solid material is present, it can produce strong diffraction rings
that interfere with the interpretation of the signal from the sample.
If diffraction is not a diagnostic, it is less important to melt the shield layers.
In some respects, strength in the solid x-ray shield may be helpful in suppressing Rayleigh-Taylor instability growth,
for instance from the ablation process.
Material properties experiments at the National Ignition Facility can reach
pressures exceeding 1\,TPa, so the strength is bound to vary greatly from ambient properties.
We have not found a strength model for Re for use at elevated pressure. 

\section{Steinberg-Guinan model}
In high-pressure studies and applications,
strength is typically manifested as a correction to a mechanical response
dominated by the scalar equation of state.
Plasticity is inherently a process of kinetic relaxation of shear stresses,
involving the motion and evolution of defects (dislocations and disclinations) in the crystal lattice.
However, yield-based strength models are commonly used as they are simpler to implement in hydrocodes
and are often sufficiently accurate despite neglecting the underlying processes of plasticity.

We chose the Steinberg-Guinan (SG) yield model \cite{Guinan1974} for its wide availability in hydrocodes and extensive use in modeling strength under dynamic loading.
In the SG model, the shear modulus $G$ is taken to vary as
\begin{equation}
G=G_0\left[1+Ap/\eta^{1/3}-B(T-T_0)\right]\quad:\quad\eta\equiv\rho/\rho_0
\label{eq:g}
\end{equation}
where $p$ is the pressure, $\rho$ the mass density, $T$ the temperature,
$A$ and $B$ pressure-hardening and thermal-softening parameters,
and subscript 0 denotes ambient values.

In the SG model, the flow stress $Y$ is taken to vary as 
\begin{equation}
Y = Y_0 f(\epsilon) G/G_0
\end{equation}
where $\epsilon$ is the equivalent plastic strain and
\begin{equation}
f(\epsilon)=\left(1+\beta\epsilon\right)^n
\end{equation}
is a work-hardening function.
$f(\epsilon)$ is limited to a maximum value $Y_{\text{max}}/Y_0$.

\section{Shear modulus and pressure hardening}
\begin{figure}
\begin{center}\includegraphics[scale=0.72]{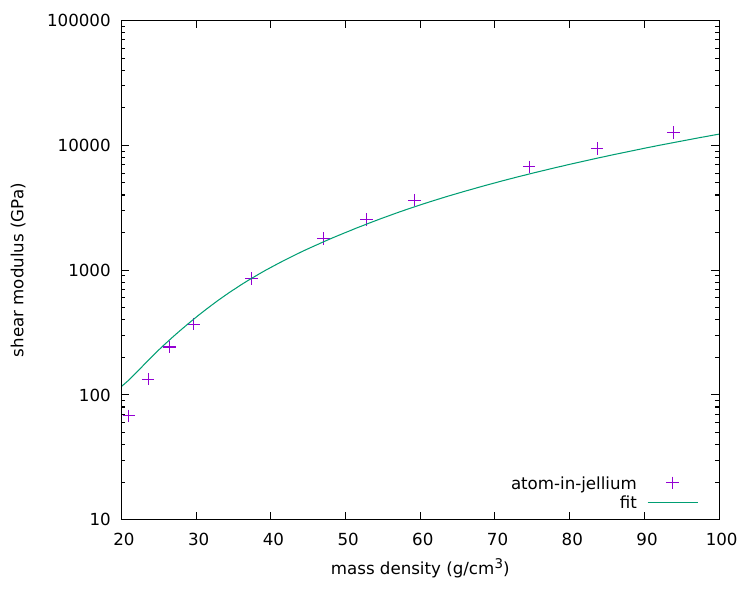}\end{center}
\caption{Fit to atom-in-jellium shear modulus predictions,
   as a function of mass density.}
\label{fig:gfit}
\end{figure}

\begin{figure}
\begin{center}\includegraphics[scale=0.72]{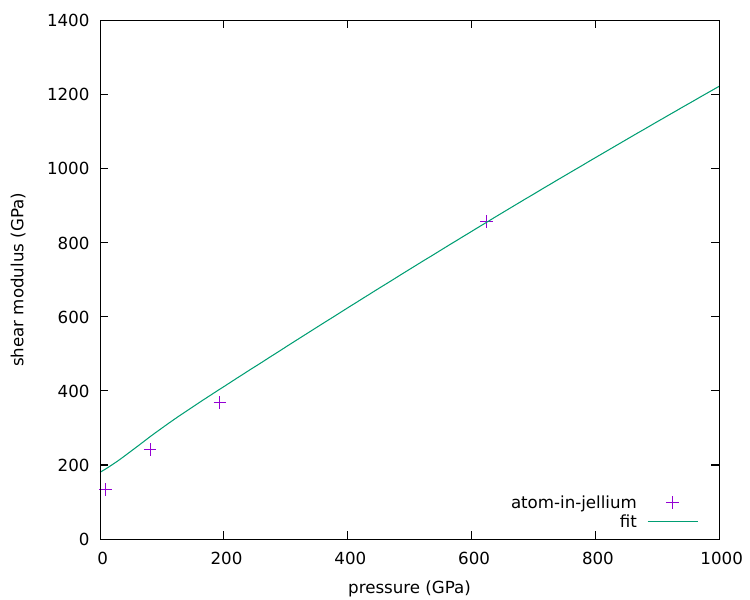}\end{center}
\caption{Fit to atom-in-jellium shear modulus predictions,
   as a function of pressure.}
\label{fig:gfitp}
\end{figure}

We recently used atom-in-jellium theory to predict the variation of shear modulus with compression to high pressures of a range of elements including Re \cite{Swift2022}.
The shear modulus was calculated in tabular form and wide-range analytic fits found.
For use with the SG model, we refitted the tabulated values using Eq.~\ref{eq:g}.
Electronic structure calculations based on density functional theory
are not perfectly accurate, the accuracy appearing worst in fractional terms
at low pressures.
Atom-in-jellium theory makes additional approximations, making the inaccuracy worse, though allowing much more efficient calculations over a wider range of states.
The predicted shear modulus at ambient was lower than observed,
so $G_0$ was taken to be the measured value of 182\,GPa and the best-fitting 
value of $A$ was found,
limiting the atom-in-jellium points to pressures below 1\,TPa
(Table~\ref{tab:sgparams}).
The fit reproduced the trend of the predictions but with a discrepancy
up to a few tens of percent (Figs~\ref{fig:gfit} and \ref{fig:gfitp}).
At this level of agreement, the fit was still reasonable to over 100\,g/cm$^3$
or over 25\,TPa.

The shear modulus depends on the texture of the material, and generally
evolves during loading.
The range can be computed from the elastic moduli, which have been measured for Re at ambient pressure and temperatures from 4 to 298\,K \cite{Shepard1965}.
It would be worth performing more detailed electronic structure predictions
of the elastic and shear moduli.

\section{Thermal softening}
For a harmonic solid, the shear modulus varies with compression but not with temperature.
Temperature variation is caused by anharmonicity, 
which is often a relatively small correction to harmonic predictions,
and at constant compression generally acts to stiffen the mechanical response.
The SG model represents the effect of compression on the shear modulus
primarily through the pressure.
At constant pressure, heating leads to thermal expansion.
The dominant effect on shear modulus is softening, because of the
reduced density rather than the increased temperature.

We can estimate the thermal softening parameter needed to exactly counteract
the thermal expansion at ambient pressure:
\begin{equation}
B=3\alpha A K
\end{equation}
where $\alpha$ is the linear thermal expansivity 
($5.61\times 10^{-6}$\,/K \cite{Touloukian1974})
and $K$ the bulk modulus 
(363.5\,GPa, computed from the elastic constants at ambient \cite{Shepard1965})
(Table~\ref{tab:sgparams}).
This value should be regarded as an upper bound on $B$ 
as anharmonicity is likely to lead to some amount of stiffening with temperature at constant compression.
However, the thermal softening term in the SG model is essentially
a first-order Taylor expansion and does not capture any variation of the
softening behavior with compression.

\section{Work-hardening}
\begin{figure}
\begin{center}\includegraphics[scale=0.72]{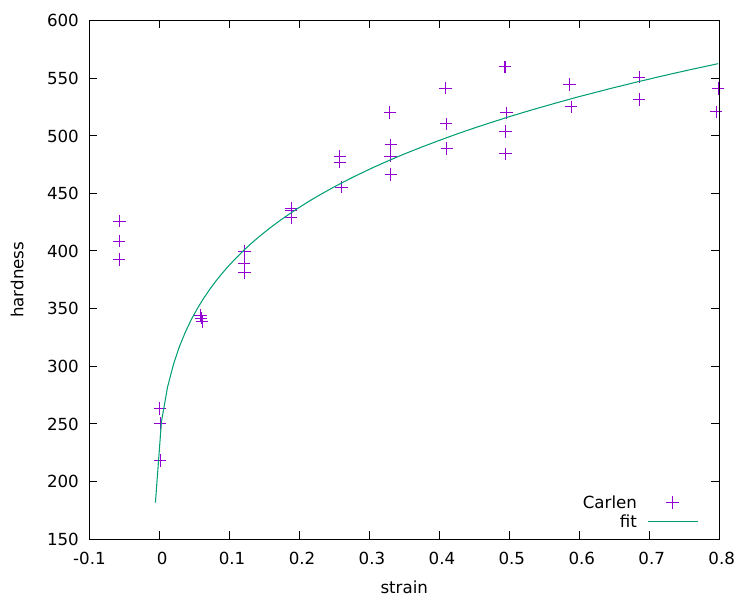}\end{center}
\caption{Fit to work-hardening data \cite{Carlen1992}.
   The points plotted at negative strain are for as-received Re before annealing,
   not used in the fit.}
\label{fig:whfit}
\end{figure}

Re exhibits significant work-hardening.
We estimated the SG hardening parameters from Knoops indentation data
on Re bars rolled to different thicknesses
\cite{Carlen1992}.
For a thickness reduced by a factor $f$, the plastic strain is
\begin{equation}
\epsilon=\frac{2}{\sqrt{3}}\ln\left(\frac 1f\right).
\end{equation}
The Knoops microhardness $h$ is approximately proportional to flow stress,
so we fitted the variation of $h/h_0$ with $\epsilon$,
where $h_0$ is the value for annealed Re before rolling
(Table~\ref{tab:sgparams}).
The SG work-hardening function reproduced the Re data reasonably well
(Fig.~\ref{fig:whfit}).
The maximum hardening $h_{\text{max}}/h_0\simeq 2.3$,
which we assumed was equal to $Y_{\text{max}}/Y_0$ and hence determined $Y_{\text{max}}$.

\section{Flow stress}
A typical flow stress for Re at ambient conditions is 0.29\,GPa.
On nanosecond timescales typical of laser-driven materials experiments,
the flow stress is typically around five times higher, so we took
$Y_0=1.45$\,GPa.

In the SG model, pressure hardening of the flow stress is assumed to
be the same as for the shear modulus.
Constraints on the flow stress can be deduced from
the radial strain variation in a sample compressed in a diamond anvil cell
without using a hydrostatic medium.
Measurements for Re suggest that $Y$ may reach 10-15\,GPa 
at pressures of 100-120\,GPa \cite{Jeanloz1991}, though the measurements
have a large uncertainty.
The SG pressure-hardening parameters deduced from the shear modulus predictions
give $Y\simeq 2.8$\,GPa at $p=120$\,GPa.
A variant of the SG model includes a separate pressure-hardening parameter
for the flow stress, or a multiplier for the shear modulus hardening.
To be consistent with the diamond anvil data,
this multiplier could be in the range 4-6.
More detailed studies of the shear modulus and flow stress would shed more light,
as might a dislocation-based model of plastic flow.

\begin{table*}
\caption{Steinberg-Guinan strength parameters for Re.}
\label{tab:sgparams}
\begin{center}
\begin{tabular}{|l|l|l|l|}\hline
$G_0$ & 182 & GPa & Depends on texture, may evolve during loading. \\
$A$ & $(7.4\pm 0.6)\times 10^{-3}$ & 1/GPa & Atom-in-jellium fit, valid to $> 25$\,TPa. \\
$B$ & $4.516\times 10^{-5}$ & 1/K & Upper bound. \\
$Y_0$ & 1.45 & GPa & Estimate for nanosecond scale. \\
$\beta$ & $132\pm 68$ & & Fit to ambient hardening data. \\
$n$ & $0.185\pm 0.015$ & & '' \\
$Y_{\text{max}}$ & 3.35 & GPa & Ratio from $Y_0$ fitted to ambient hardening data. \\
\hline\end{tabular}
\end{center}
\end{table*}

\section{Conclusions}
We determined parameters in the SG strength model for Re,
using our recent theoretical prediction of the variation of shear modulus with compression
and previous measurements of work-hardening in rolled bars.
The SG model of the shear modulus, fitted to predictions up to 1\,TPa, 
was still in reasonable agreement at pressures exceeding 25\,TPa.
The predicted effect of pressure-hardening on the flow stress
was significantly less than suggested by previous measurements of 
radial strain variation during compression in a diamond anvil cell,
suggesting that the strength may be several times higher at megabar pressures.

\section{Acknowledgments}
This work was performed under the auspices of the
U.S. Department of Energy by Lawrence Livermore National Laboratory
under Contract DE-AC52-07NA27344.

\end{document}